# NEW RESULTS FROM THE AUGER OBSERVATORY

GIORGIO MATTHIAE
on behalf of the Pierre Auger Collaboration
*University and Sezione INFN of Roma Tor Vergata, Roma, Italy*

ABSTRACT

The Auger project was designed to study the high-energy cosmic rays by measuring the properties of the showers produced in the atmosphere. The Southern Auger Observatory has taken data since January 2004 and is now completed. Results on mass composition, energy spectrum and anisotropy of the arrival directions are presented together with upper limits on the neutrino fraction. The most important result is the recent observation of correlations with nearby extragalactic objects.

## 1. Introduction

It has been known for a long time that the flux of cosmic rays decreases with the primary energy E following approximately a power law $E^{-\gamma}$ with spectral index $\gamma$ roughly equal to 3. A compilation [1] of the flux as measured until the year 2000 is shown in Fig.1.

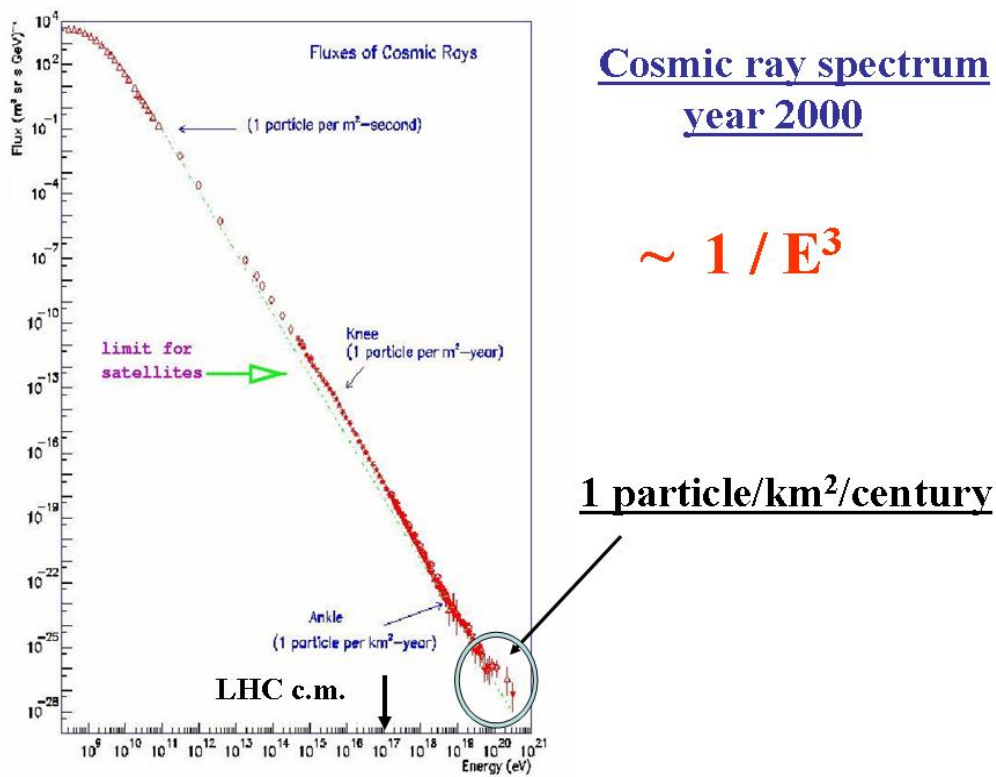

Figure 1.   The flux of primary cosmic rays as a function of energy.  Data until the year 2000.

The spectrum exhibits interesting features, usually called the "knee" and the "ankle". At the energy of the "knee"  (~ $3 \times 10^{15}$ eV) the spectral index changes from approximately 2.7 to 3.1.  Another change of the spectral index is observed in the region of the "ankle", around a few $10^{18}$ eV.   In the region above $10^{19}$ eV the flux of the primaries is extremely low, of  the order of     1 particle/ km$^2$/ century.  Therefore the study of cosmic rays in this very high-energy region requires detectors with very large acceptance.

A recent compilation [2] which presents the product (Flux x $E^{2.5}$ ) as a function of the primary energy is shown in Fig.2.

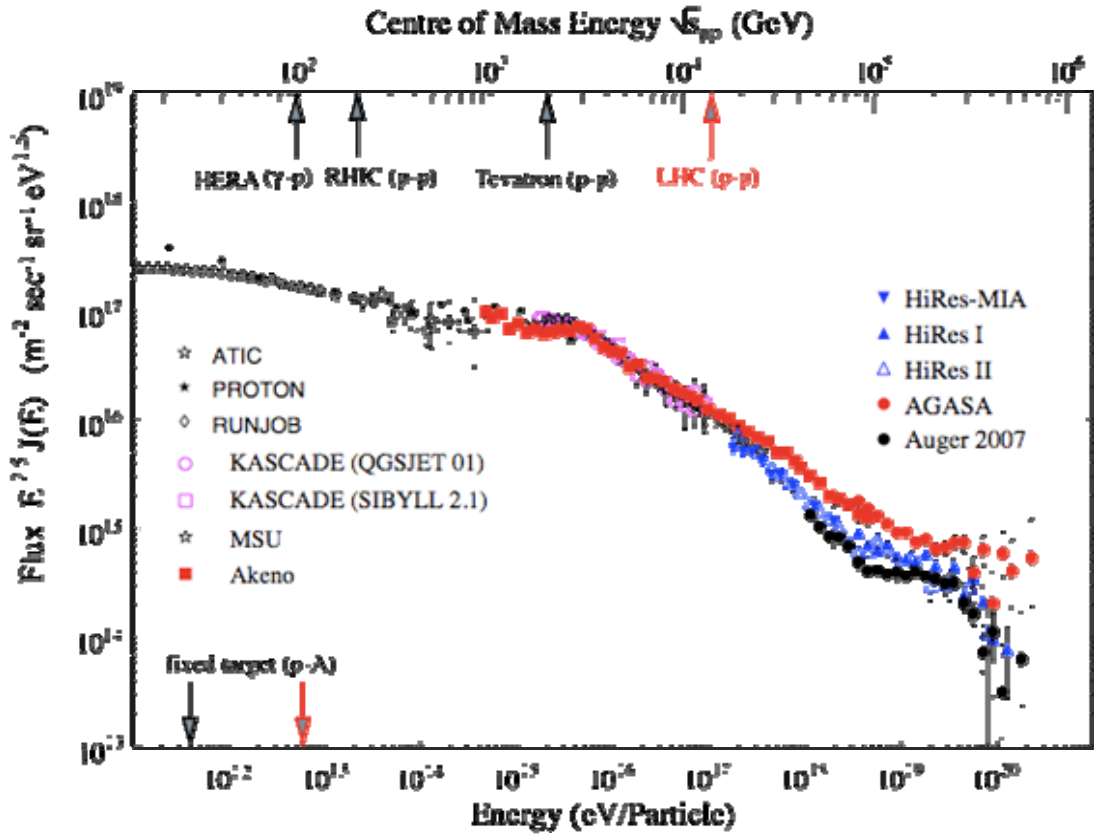

Figure 2.   Recent compilation of the flux of primary cosmic rays as a function of energy.  On the vertical axis the product of the flux times the power of the energy $E^{2.5}$ .

The Auger Observatory is dedicated to the study of the region at the very end of the spectrum.  The two features present in this region are the "ankle"  and a fast decrease of the  flux above  ~ $4 \times 10^{19}$ eV  which is usually attributed to the Greisen, Zatsepin and Kuz'min effect (GZK) i.e. to the interaction of the primaries with the Cosmic Microwave Background  (CMB).

## 2. The Auger Observatory

Two Observatories, one in the Northern and one in the Southern hemisphere are foreseen in the Auger project, to achieve a full exploration of the sky. The Southern Auger Observatory [3] is located in the *"Pampa Amarilla"* , near the small town of Malargüe in the province of Mendoza (Argentina) at the latitude of about $35^0$ S and altitude of 1400 above sea level. The region is flat, with very low population density and favorable atmospheric conditions. The Observatory is a hybrid system, a combination of a large surface array and a fluorescence detector.

The surface detector (SD) is a large array of 1600 water Cherenkov units spaced at a distance of 1.5 km and covering a total area of 3000 km$^2$. Each SD unit is a plastic tank of cylindrical shape with size 10 m$^2$ x 1.2 m filled with purified water. Technical details are given in Fig. 3. The surface detector measures the front of the shower as it reaches ground. The surface detector units, which are activated by the event, record the particle density and the time of arrival.

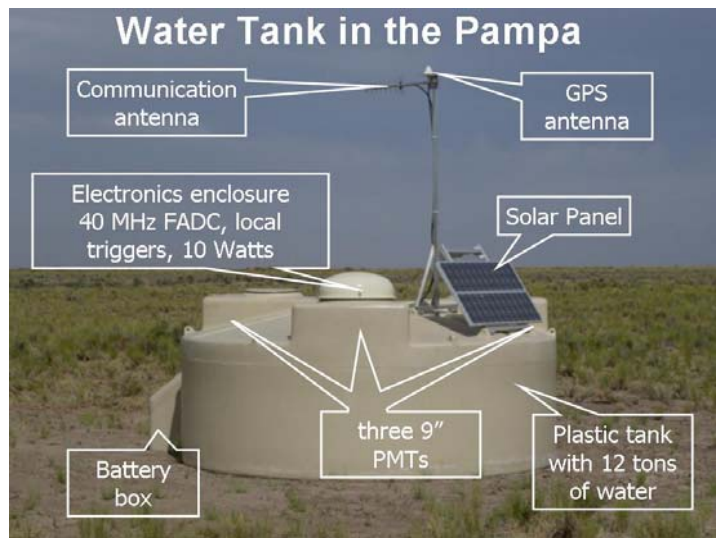

Figure 3. Picture of a water Cherenkov unit (tank) of the Surface Detector of the Auger Observatory. The insets give explanations on the various components of the system.

The fluorescence detector (FD) consists of 24 telescopes located in four stations which are built on the top of small elevations on the perimeter of the site. The telescopes measure the shower development in the air by observing the fluorescence light. Each telescope (see Fig.4) has a 12 m$^2$ spherical mirror with curvature radius of 3.4 m and a camera with 440 photomultipliers.

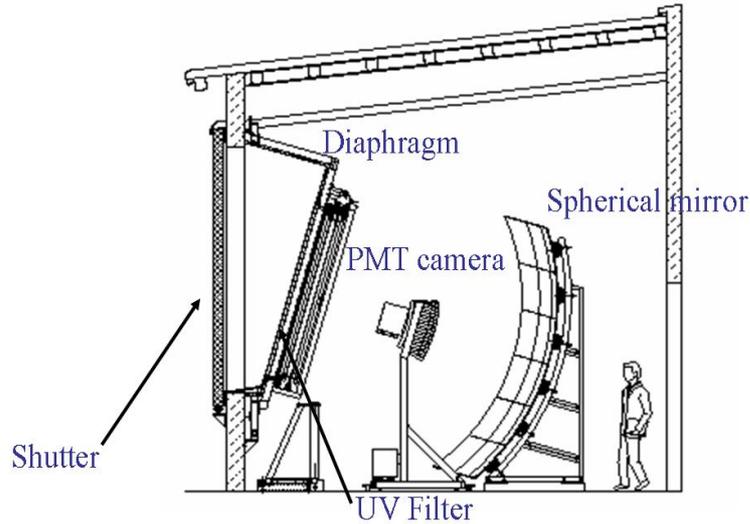

Figure 4. Sketch of a fluorescence telescope. The various components are indicated.

The field of view of each telescope is $30^0$ x $30^0$. UV filters placed on the diaphragm reject light outside the 300-400 nm spectrum of the air fluorescence. The FD may operate only in clear moonless nights and therefore with a duty cycle of about 12%.

Attenuation of the fluorescence light due to Rayleigh and aerosol scattering along the path from the shower to the telescope is measured systematically with atmospheric monitors including LIDAR systems.

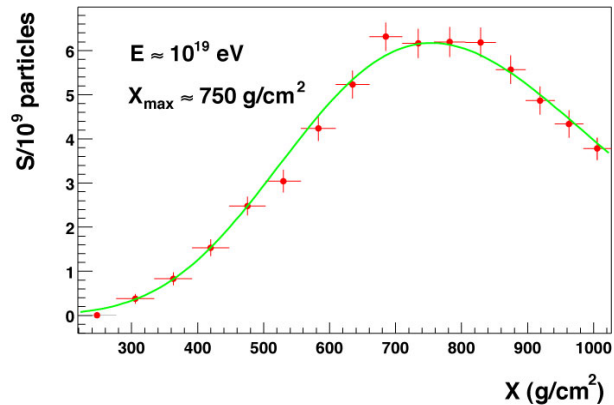

Figure 5. Example of a measured longitudinal profile of a high-energy shower.

An example of a longitudinal profile of a shower as measured by the FD is shown in Fig. 5 where the number of particles of the shower is plotted as a function of the atmospheric depth. In order to obtain the shower profile, the contamination due to Cherenkov light has to be subtracted. The empirical formula by Gaisser and Hillas is used to fit the data.

An example of an event of very high energy as observed by the SD is shown in Fig. 6.

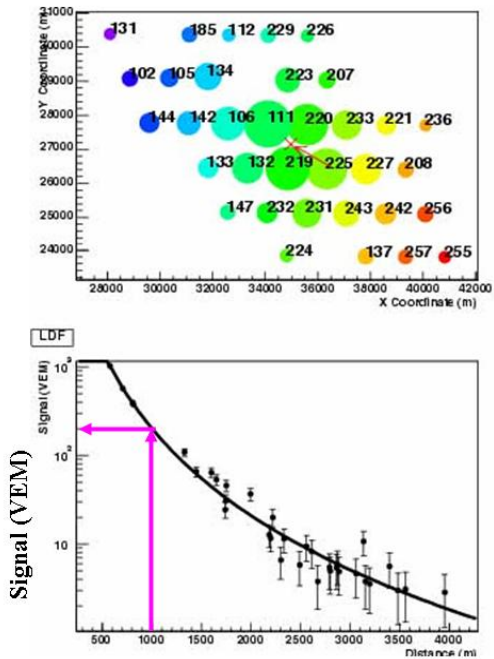

Figure 6. Example of a very high energy event as observed by the SD. The shower has activated 34 units (tanks) of the surface detector distributed over an area of more than 50 km$^2$

The signals of the surface detector are expressed in units of Vertical Equivalent Muons (VEM) which represents the signal produced by a muon traversing the tank vertically. The flux of cosmic ray muons provides a continuous monitoring of the SD. From the magnitude and the time of the observed signals for all activated SD units, one derives the direction of the axis of the shower and the point of impact at ground. The left bottom panel of Fig. 6 shows the signal, expressed in units of VEM as a function of the distance from the shower axis.

A simple analytical expression known as Lateral Distribution Function (LDF) is then fitted to the data to obtain the signal at the distance of 1000 m from the axis. From model calculations it is expected that the interpolated signal at some fixed optimal distance from the shower core, S(1000) for the SD array of the Auger Observatory, is a good energy estimator in the sense that it is well correlated with the energy of the primary [4].

## 3. Mass composition
The direct method to study the mass composition is based on the measurement of the longitudinal profile of the showers. It is well known that for a given energy protons are more penetrating than light/medium nuclei which interact essentially as a collection of nucleons.

The depth of the maximum of the shower profile $X_{max}$, as measured by the fluorescence telescopes, is well correlated with the particle mass. The principle of the method is indicated in Fig. 7. The FD detector of Auger can measure $X_{max}$ with systematic uncertainty of about 15 g/cm$^2$.

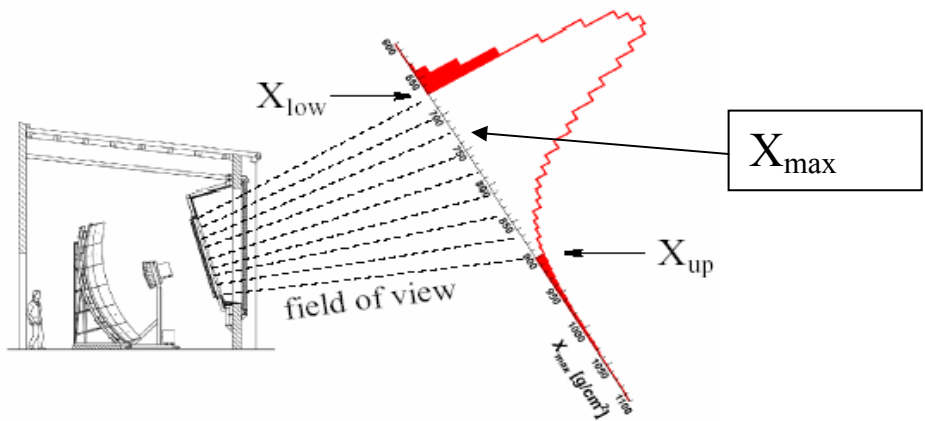

Figure 7. Illustration of the measurement of the quantity $X_{max}$ by a fluorescence telescope of the Auger Observatory.

A compilation of earlier data on $X_{max}$ for energies above $10^{14}$ eV is shown in Fig. 8 where expectations from simulation programs are also given for Fe nuclei, protons and photons. The value of $X_{max}$ for protons is about 100 g cm$^{-2}$ larger than for iron.

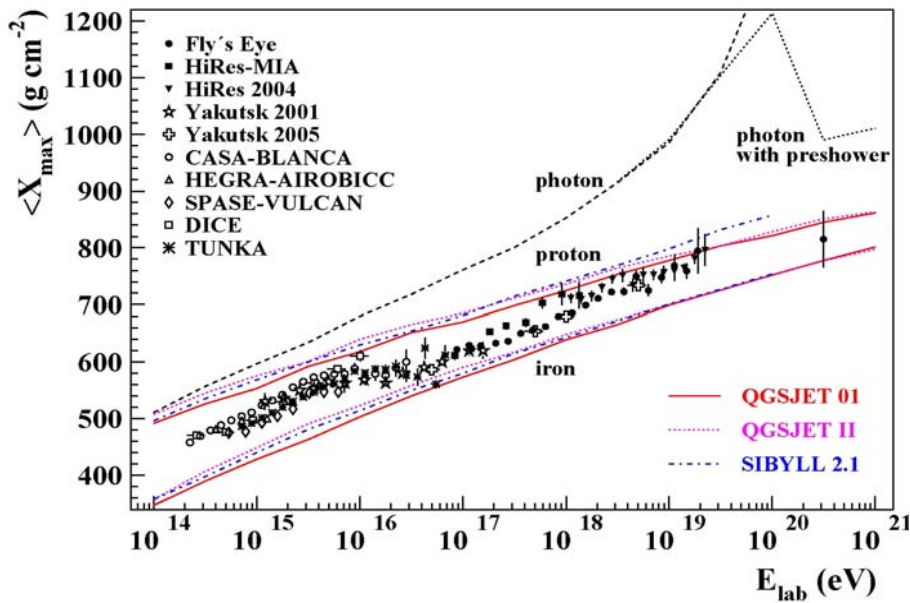

Figure 8. Compilation of earlier data on the quantity $X_{max}$ as a function of energy. Prediction of various simulation programs for incident photons, protons and iron nuclei are also shown.

Recent data from the Auger Observatory [5] are presented in Fig. 9 together with the predictions of various simulation programs. In spite of the still low statistics, the data indicate some change of regime around 2 EeV where the slope (elongation rate) changes. At the highest energies the trend is intermediate between protons and Fe nuclei with a mean mass number of about 5.

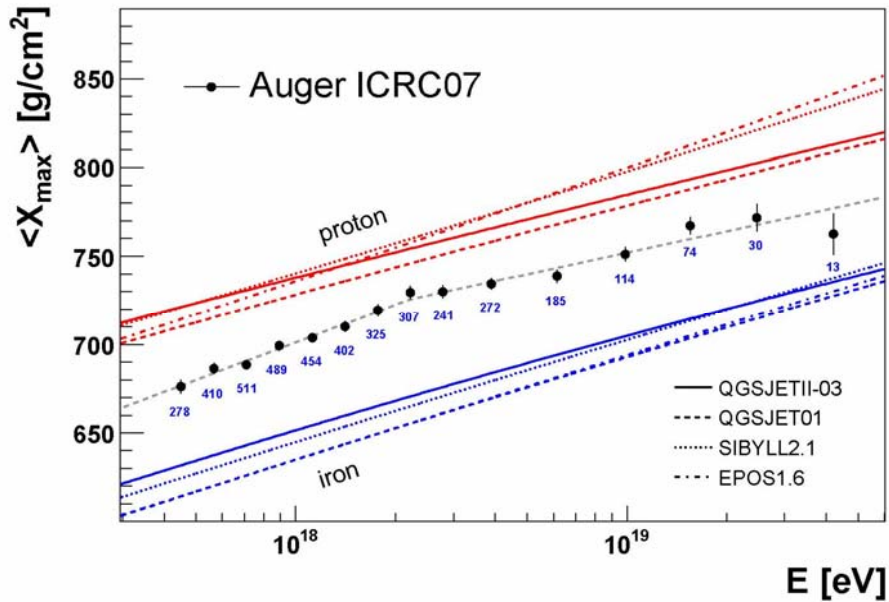

Figure 9. The Auger data on the quantity $X_{max}$ are plotted as a function of energy and compared to predictions of simulation programs for protons and iron nuclei. The number of events for each data point is also shown. The errors shown are statistical.

## 4. Search for tau neutrinos

All models of high-energy cosmic rays predict neutrino fluxes from the decay of charged pions which are produced either in interactions of the cosmic rays in the source or in interactions with background radiation as the CMB (GZK neutrinos). Neutrino oscillations during propagation in space will produce flavour mixing and give rise to tau neutrinos.

The curvature of the Earth provides a way of detecting tau neutrinos. They may enter the Earth just below the horizon and make a charged current interaction that produces a tau lepton which will decay in flight and initiate a nearly horizontal shower (Earth skimming effect). From simulation of such showers one expects to observe a very special configuration in the surface detector, quite different from the usual hadron initiated showers.

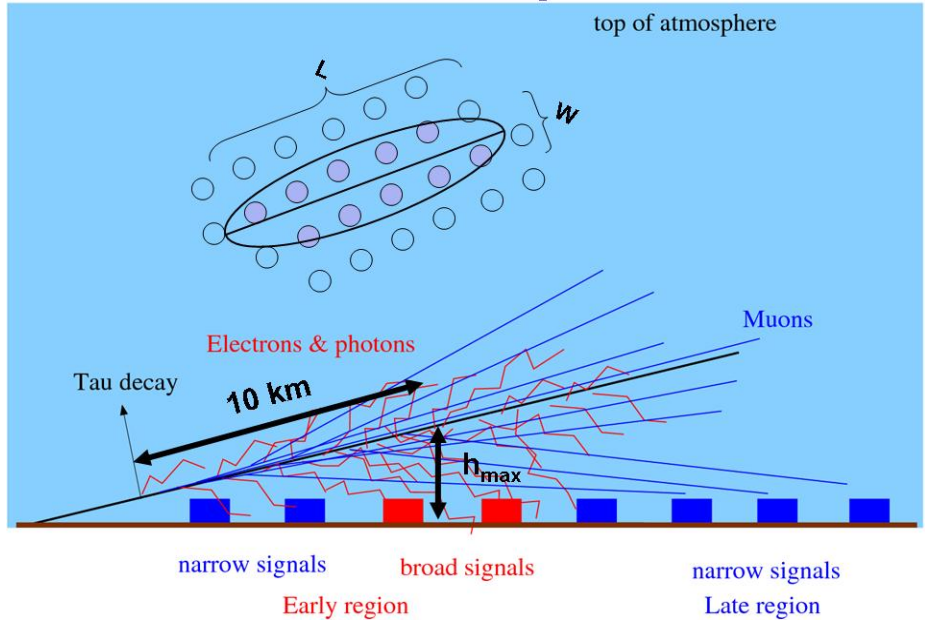

Figure 10. Sketch of how a typical shower induced by a tau neutrino may appear. A very inclined shower coming from below the horizon activates a region of the surface detector with elongated shape. The time sequence of the signals from the surface detector units is also typical.

No neutrino candidate was found and therefore at present we can only derive the upper limit [6] shown in Fig. 11.

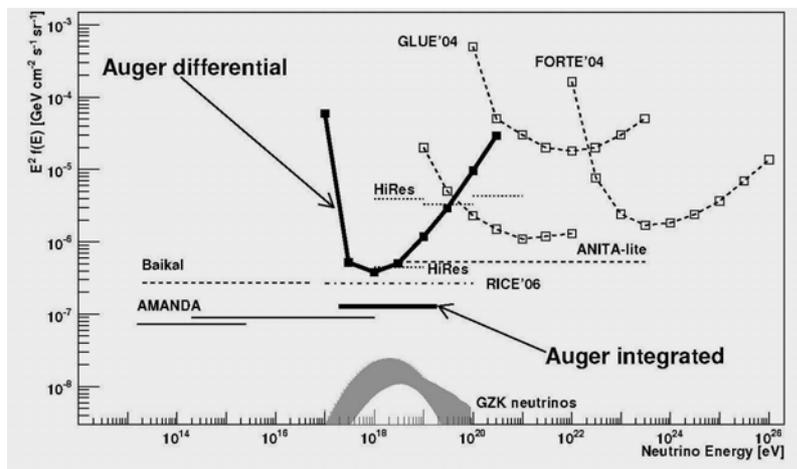

Figure 11. Upper limit on the tau neutrino flux from Auger and other experiments. The neutrino flux expected from GZK events is also shown.

## 5. The energy spectrum

As already noted in the Introduction, an important feature of the spectrum in the energy region above ~$10^{19}$ eV is the mechanism suggested by Greisen, Zatsepin and Kuz'min which is known as GZK effect. It is due to the interactions of the cosmic rays with the low energy photons of the Cosmic Microwave Background.

Protons with energy above the threshold for photoproduction of pions (~ $4 \times 10^{19}$ eV) will lose energy as they travel in space. The value of the energy where an integral power-law spectrum would be reduced to one half is $5.3 \times 10^{19}$ eV [7]. The energy loss per interaction is about 15 – 20 %.

This leads to the concept of horizon. Protons of very high energy cannot come from too far away. At ~ $5 \times 10^{19}$ eV most of the observed particles must have come from sources within about 100 Mpc.

Production of electron-positron pairs is also present but it is less effective than photo-pion production. However, this process is predicted [7] to be responsible for a feature related to the so-called "ankle", a shallow minimum (or "dip") in the plot of the flux times $E^3$ which is centered at energies of a few $10^{18}$ eV.

For nuclei, in addition to pion photoproduction, nuclear photodissociation processes have to be taken into account as (γ, n), (γ, p) etc.

In the past there was a controversy on the actual presence of the GZK suppression. The AGASA data did not show a suppression, contrary to the preliminary data of HiRes. The experimental situation is now clarified by the final data of HiRes [8], shown in Fig. 12 and by the data of Auger (see also compilation of Fig.2). The HiRes data clearly show a steepening of the spectrum above $10^{19.6}$ eV with a fitted value of the spectral index $\gamma = 5.1 \pm 0.7$. The steepening agrees with the expectations from the GZK mechanism.

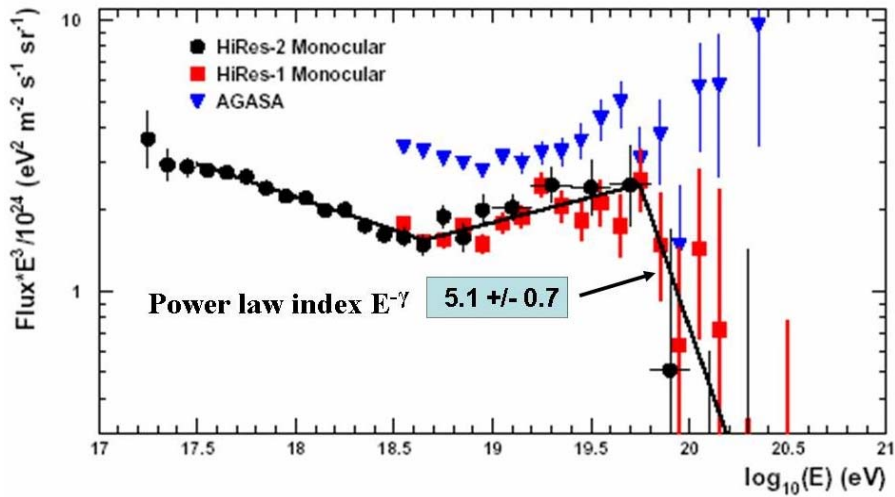

**Figure 12.** The final HiRes results on the energy spectrum are presented as (Flux x $E^3$) and compared to the earlier AGASA data. The GZK suppression is clearly seen. In addition the shallow minimum centered around $10^{18.6}$ eV is also evident.

The method used by Auger to measure the energy spectrum exploits the hybrid nature of the experiment with the aim of using the data itself rather than simulations.

For each event, the energy estimator S(1000) is obtained as discussed in Section 2. A correction to the energy estimator S(1000) depending on the zenith angle is needed because the effective atmosphere thickness seen by showers before reaching ground changes with the zenith angle. The value of S(1000) corresponding to the median zenith angle of $38^0$ is used as reference and the zenith angle dependence of the energy estimator is determined assuming that the arrival directions are isotropically distributed. This procedure is traditionally called "Constant intensity cut method".

The absolute calibration of S(1000) is derived from the hybrid events using the calorimetric energy measured by the FD which is then corrected for the missing energy (neutrinos and muons) using the mean value between proton and iron (10% correction at $10^{19}$ eV with uncertainty ± 2%). This absolute calibration, which defines the energy scale, is at present affected by a systematic error of about ± 20%, mainly due to uncertainties on the fluorescence yield and on the calibration of the FD telescopes.
The energy calibration, obtained from the subset of hybrid events (see Fig.13) is then used for the full set of events with higher statistics as measured by the SD.

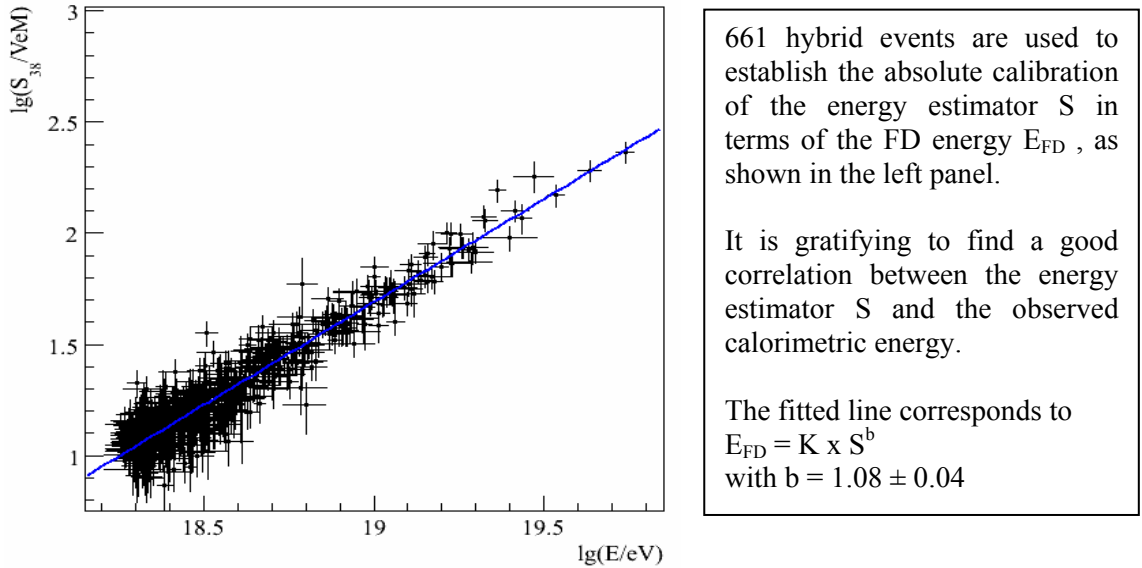

661 hybrid events are used to establish the absolute calibration of the energy estimator S in terms of the FD energy $E_{FD}$, as shown in the left panel.

It is gratifying to find a good correlation between the energy estimator S and the observed calorimetric energy.

The fitted line corresponds to
$E_{FD} = K \times S^b$
with $b = 1.08 \pm 0.04$

Figure 13. Calibration of the energy estimator S(1000) using the calorimetric energy from the FD.

The energy spectrum measured by the Auger surface array for zenith angles less than $60^0$ [9] is shown in Fig. 14. The data refer to energies above $3 \times 10^{18}$ eV where the trigger is fully efficient.

Above $4 \times 10^{19}$ eV the spectrum shows a clear change of slope in agreement with the expectations from the GZK effect. A simple way to describe this feature consists of fitting a power law in the energy region up to $4 \times 10^{19}$ eV and then to extrapolate this form to higher energies. The number of observed events is much less than expected from this extrapolation. For energies above $4 \times 10^{19}$ eV, we observe 69 events while the extrapolation gives 167±3, and above $10^{20}$ eV we observe 1 event while we would expect 35±1 from the extrapolation.

A better way of analyzing the shape of the energy spectrum is by taking the relative difference of the data with respect to the reference form $J_s = A\ E^{-2.69}$. The result is presented in Fig.15.

The change of the spectral index γ at the ankle and on the region of the GZK effect are clearly visible. Numerical values of the spectral index γ in the two different energy intervals are given in Table 1.

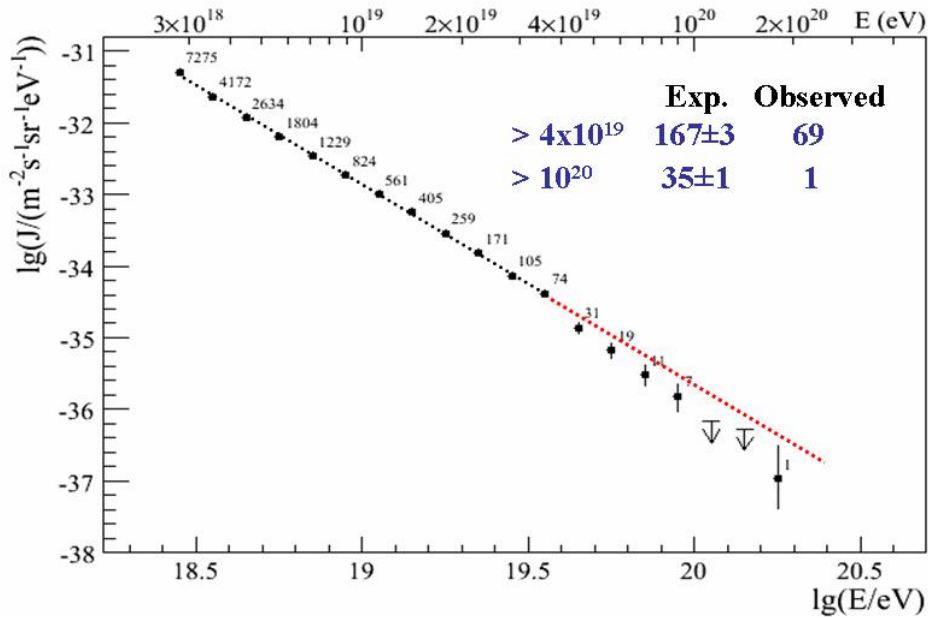

Figure 14. The Auger energy spectrum from the SD with energy calibration from the FD.

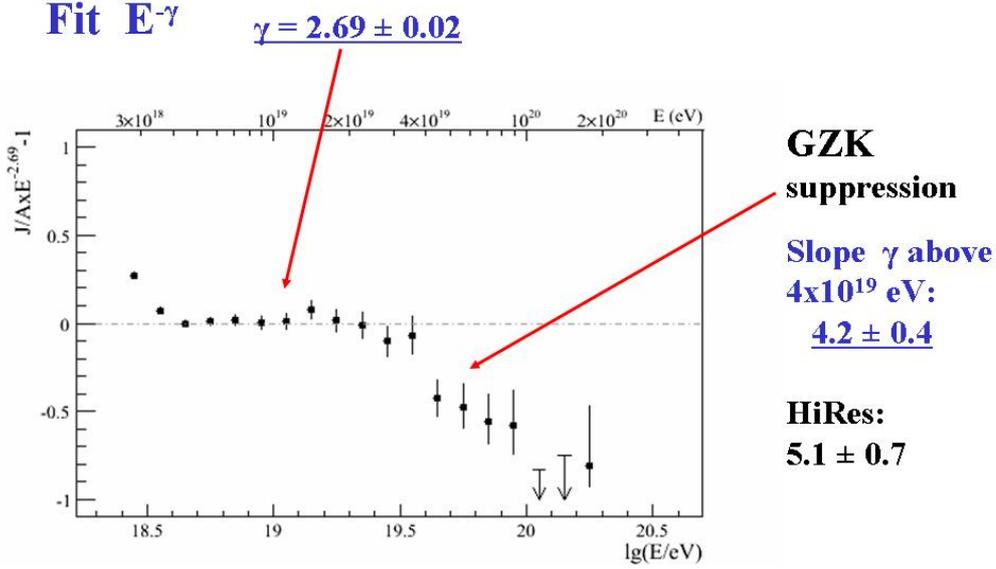

Figure 15. The Auger spectrum is presented as relative difference with respect to the form $J_s = A\,E^{-2.69}$ which describes the data well between the ankle and the beginning of the GZK suppression.

Table 1.  Numerical values of the spectral index γ of the power law fits in the two energy intervals. The Auger results have an additional systematic error of 0.06.

|  | Auger | HiRes |
| --- | --- | --- |
| γ for ($E_{ankle} < E < E_{GZK}$) | 2.69 ± 0.02 | 2.81 ± 0.03 |
| γ for ($E > E_{GZK}$) | 4.2 ± 0.4 | 5.1 ± 0.7 |

## 6. Anisotropy studies

In the study of anisotropy the Auger Observatory may exploit the good angular resolution of the SD which is better than one degree at high energy.
Observation of an excess from the region of the Galactic centre at the level of 4.5 σ, in the energy region 1.0 – 2.5 EeV and with angular scale of $20^0$, was reported by AGASA [10]. The Auger Observatory is suitable for this study because the Galactic centre (constellation of Sagittarius), lies well in the field of view of the experiment. However, at present the Auger data [11] don't confirm the AGASA result.

The Auger collaboration has done an extensive search for correlation of the high-energy events with known astrophysical objects. This study started early in 2004 and the results from data collected until August 2007 have been published recently [12].
The arrival direction of high-energy events was compared to the direction in the sky of the galaxies with active nucleus (AGN)[13].

A sophisticated analysis described in ref. 12 has shown that a clear correlation, within an angle ψ about equal to 3 degrees, exists between the arrival directions of cosmic rays with energy above about 60 EeV and AGNs at distances less than about 75 Mpc. The Véron-Cetty / Véron catalog was used. The direction on the sky of the events and of the AGNs is shown in galactic coordinates in Fig. 15. Out of 27 events, 20 correlate with AGNs. Two events are correlated within less than 3 degrees with Cen A, a strong radio source at the distance of about 4 Mpc.

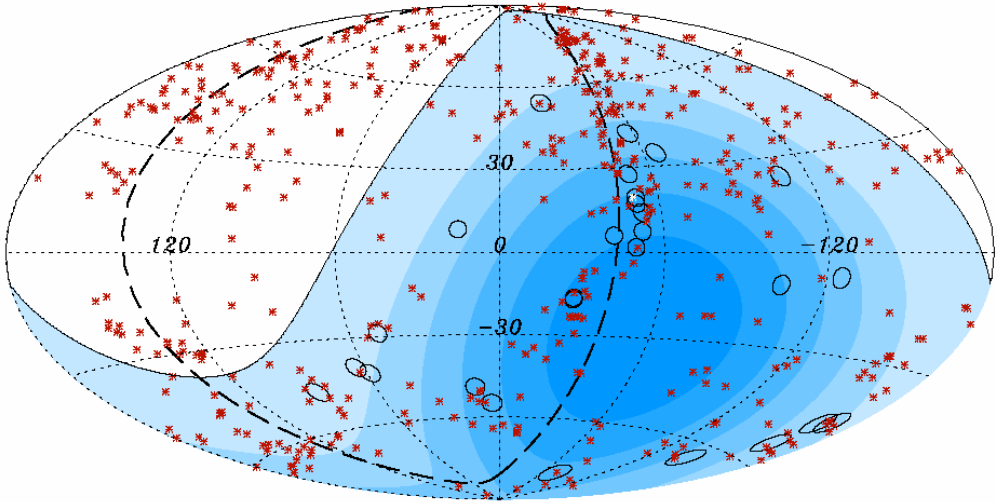

Figure 15. Plot in galactic coordinates showing the events with energy larger than 57 EeV as small circles of radius 3.2 degrees. The supergalactic plane is shown as a dashed line. The red crosses indicate the position of AGN within 71 Mpc. Cen A, one of the nearest AGN is marked in white. The white region of the sky is not accessible from the Southern Auger Observatory. Darker blue regions indicate larger relative exposure.

The results are summarized in Table 2. The first exploratory analysis has shown that 12 out of 15 events with energy above 57 EeV were correlated with AGN at distances less than 75 Mpc, within 3.1 degrees while only 3.2 were expected to be correlated by chance for an isotropic distribution.

As a consequence of this result, a prescribed test was defined to see whether the isotropy hypothesis had to be accepted or rejected. The same set of parameters and the same reconstruction algorithms were used. The second independent set (see Table 2, row #2) satisfied the test and the probability for this single configuration to happen by chance if the flux was isotropic is $1.7 \times 10^{-3}$.

A complete reanalysis of the data set gave the results reported in Table 2, row#3. Out of 27 events, 20 were found to correlate with a chance probability of the order of $10^{-5}$.

The correlation becomes statistically more significant if the events in the band around the galactic plane (latitude |b| <12 degrees) are removed. For this subset of 21 events, 19 are correlated with AGN. Elimination of the galactic plane region is motivated by the incompleteness of the catalog in this region and by the expected stronger effect of the galactic magnetic field which is known to be concentrated in the galactic disk.

**Table 2**. Results of the analysis for the first set, the second independent set, the reanalysis of the full set and for the full data set excluding the galactic plane region are reported.

| | Number of events E >57 EeV | Events correlated with AGN $\psi$ = 3.1 degree | Events expected for isotropy |
|---|---|---|---|
| Exploratory scan 1 Jan 04- 27 May 06 | 15 | 12 | 3.2 |
| Second independent set 27 May 06–31 Aug 07 | 13 | 8 | 2.7 |
| Full data set (about 1.2 year full Auger) | 27 | 20 | 5.6 |
| Full data set excluding galactic plane region | 21 | 19 | 5.0 |

The distribution of the separation angle between the direction of the 27 high-energy events and the nearest AGN is shown in Fig. 16. For comparison the histogram expected for isotropic distribution of the events is also shown. The data are clearly not consistent with an isotropic distribution.

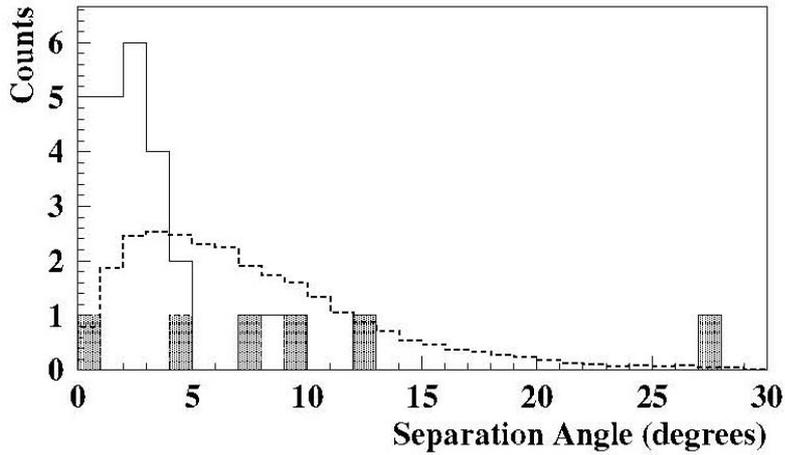

Figure 16. Distribution of the angle between each event of energy larger than 57 EeV and the nearest AGN. The dotted histogram represents the expectation for isotropic distribution. The histogram shows the data while the 6 shaded areas represent the events removed because close to the galactic plane.